\documentclass[12pt,draftclsnofoot,onecolumn]{IEEEtran}
\usepackage{cite}
\usepackage{amsmath,amssymb,amsfonts}
\usepackage{algorithm}
\usepackage{algorithmic}
\usepackage{CJK}
\usepackage{longtable,booktabs}
\usepackage{booktabs}
\usepackage{graphicx}
\usepackage{subfigure}
\usepackage{textcomp}
\usepackage{flushend}
\usepackage{amsthm}
\usepackage{upgreek}
\usepackage{graphicx}
\usepackage{supertabular}
\usepackage[bottom]{footmisc}
\usepackage{stfloats}
\usepackage{color}
\usepackage{booktabs}
\usepackage{multirow}
\usepackage{array}
\usepackage{geometry}
\usepackage{epstopdf}
\UseRawInputEncoding 

\def\BibTeX{{\rm B\kern-.05em{\sc i\kern-.025em b}\kern-.08em
    T\kern-.1667em\lower.7ex\hbox{E}\kern-.125emX}}

\newcommand\bfb{\ensuremath{{\mathbf b}}}
\newcommand\bfB{\ensuremath{{\mathbf B}}}

\newcommand\bfA{\ensuremath{{\mathbf A}}}

\newcommand\bfS{\ensuremath{{\mathbf S}}}
\newcommand\bfG{\ensuremath{{\mathbf G}}}
\newcommand\bff{\ensuremath{{\mathbf f}}}
\newcommand\bfH{\ensuremath{{\mathbf H}}}
\newcommand\bfy{\ensuremath{{\mathbf y}}}

\newcommand\bfR{\ensuremath{{\mathbf R}}}
\newcommand\bfY{\ensuremath{{\mathbf Y}}}
\newcommand\bfx{\ensuremath{{\mathbf x}}}
\newcommand\bfX{\ensuremath{{\mathbf X}}}
\newcommand\bfz{\ensuremath{{\mathbf z}}}
\newcommand\bfZ{\ensuremath{{\mathbf Z}}}
\newcommand\bfv{\ensuremath{{\mathbf v}}}
\newcommand\bfw{\ensuremath{{\mathbf w}}}
\newcommand\bfW{\ensuremath{{\mathbf W}}}
\newcommand\bfV{\ensuremath{{\mathbf V}}}
\newcommand\bfa{\ensuremath{{\mathbf a}}}
\newcommand\bfn{\ensuremath{{\mathbf n}}}
\newcommand\bfN{\ensuremath{{\mathbf N}}}
\newcommand\bfI{\ensuremath{{\mathbf I}}}

\newcommand\rmT{\ensuremath{{\mathrm T}}}

\newcommand\rmB{\ensuremath{{\mathrm B}}}
\newcommand\rmH{\ensuremath{{\mathrm H}}}
\newcommand\rmBI{\ensuremath{{\mathrm {BI}}}}
\newcommand\rmLoS{\ensuremath{{\mathrm {LoS}}}}
\newcommand\rmNLoS{\ensuremath{{\mathrm {NLoS}}}}

\newcommand\rmdiag{\ensuremath{{\mathrm {diag}}}}

\newcommand\rmS{\ensuremath{{\mathrm S}}}
\newcommand\rmI{\ensuremath{{\mathrm I}}}

\newcommand\calN{\ensuremath{{\mathcal N}}}
\newcommand\calR{\ensuremath{{\mathcal R}}}
\newcommand\calH{\ensuremath{{\mathcal H}}}
\newcommand\bbC{\ensuremath{{\mathbb C}}}

\begin{document}

\title{Deep-Learning-Based Channel Estimation \\ for IRS-Assisted ISAC System}

\author{\IEEEauthorblockN{
{Yu Liu}$^*$,
{Ibrahim Al-Nahhal}$^\ddagger$,
{Octavia A. Dobre}$^\ddagger$,
and Fanggang Wang$^*$\\} 
\IEEEauthorblockA{
$\,^*$State Key Laboratory of Rail Traffic Control and Safety, \\Beijing Jiaotong University, Beijing, 100044, China\\
$\,^\ddagger$Faculty of Engineering and Applied Science, \\Memorial University, St. John's, NL A1B 3X9, Canada}

\thanks{

{Digital Object Identifier: 10.1109/GLOBECOM48099.2022.10001672}

{This article is available at: https://ieeexplore.ieee.org/document/10001672}
}
}

\maketitle

\begin{abstract}
  Integrated sensing and \textcolor{black}{communication} (ISAC) \textcolor{black}{and} intelligent reflecting surface (IRS) are viewed as promising technologies for future generations of wireless networks.
  This paper investigates the channel estimation problem in an IRS-assisted ISAC system. 
  A \textcolor{black}{deep-learning} framework is proposed to estimate the sensing and communication (S\&C) channels in such a system.
  Considering different propagation environments of the S\&C channels, two deep neural \textcolor{black}{network} (DNN) architectures are designed to realize this framework.
  The \textcolor{black}{first DNN} is devised at the ISAC base station for \textcolor{black}{estimating} the sensing channel, while the \textcolor{black}{second DNN architecture} is \textcolor{black}{assigned} to each downlink user equipment to estimate its communication channel.
  Moreover, the input-output pairs \textcolor{black}{to train} the DNNs \textcolor{black}{are carefully designed}.
  Simulation results show the superiority of the proposed estimation approach compared to the benchmark scheme under various  \textcolor{black}{signal-to-noise ratio} conditions and \textcolor{black}{system parameters}.

\end{abstract}

\section{Introduction}\label{sec:intro}
\IEEEPARstart{T}{he} integrated sensing and {communication} (ISAC) technology has been envisioned as a \textcolor{black}{promising} candidate to enhance spectral and energy
efficiencies in \textcolor{black}{future generations of wireless networks} \cite{ref:ISAC-survey,ref:ChModel-b}. 
To \textcolor{black}{efficiently} merge sensing and communication (S\&C) functionalities into a single system, the performance \textcolor{black}{trade-off} between S\&C has recently \textcolor{black}{attracted} research attention \cite{ref:ISAC-improve-S1,ref:ISAC-improve-S2,ref:ISAC-improve-C1,ref:ISAC-improve-C2}. 
\textcolor{black}{The authors in \cite{ref:ISAC-improve-S1} and \cite{ref:ISAC-improve-S2} optimized} the sensing functionality, {such as} \textcolor{black}{the} target detection probability {and} target distance/angle estimation accuracy, while {ensured} \textcolor{black}{an} acceptable communication performance.
On the other \textcolor{black}{hand}, sensing-assisted communication, which sheds light on how sensing can be co-designed to improve \textcolor{black}{the} communication performance including sensing-assisted beam training, tracking, and prediction, \textcolor{black}{is also investigated in  \cite{ref:ISAC-improve-C1} and \cite{ref:ISAC-improve-C2}}.
\textcolor{black}{The} above literature generally \textcolor{black}{assumed} that the channel state information (CSI) is known at the receiver side and rarely \textcolor{black}{considered} the channel estimation problem for the ISAC systems. 

Intelligent reflecting surface (IRS) has been foreseen as another promising technology \textcolor{black}{for} future wireless networks\textcolor{black}{; it} enables the smart and reconfigurable wireless propagation environment \cite{ref:IRS-survey,ref:IRS-SCMA-optimize,ref:IRS-SCMA,ref:ChModel-refpower,ref:IRS-reviewer}.
IRS is a programmable planar surface composed of low-cost electromagnetic passive elements.
Particularly, based on the CSI of the surrounding wireless environment, IRS can bring outstanding beamforming gain to a communication system by coordinating the reflections of its passive elements.
As such, {estimating the channels} is essential in an IRS-assisted wireless communication system and has been widely researched in \textcolor{black}{the} existing literature\textcolor{black}{; it} can be classified into conventional model-driven \textcolor{black}{\cite{ref:IRS-turn-off} \cite{ref:DL-IRS-ChE-modelDriven}} and data-driven \textcolor{black}{deep-learning} (DL) \textcolor{black}{approaches} \cite{ref:DL-IRS-ChE-TWC,ref:DL-IRS-ChE-WCL,ref:DL-IRS-ChE-timevary}.

To further enhance the S\&C performance, ISAC and IRS have been jointly explored recently  \cite{ref:ChModel-pathloss-SJ,ref:IRS-ISAC-wavebeam1,ref:IRS-ISAC-wavebeam2}.
Considering their cooperation merits, the authors in \cite{ref:ChModel-pathloss-SJ} {developed} an IRS-assisted ISAC framework, which aims to optimize the sensing performance by jointly devising the active and passive beamforming under the constraints of the communication metrics.
Moreover, the joint schemes of waveform and passive beamforming design are studied for the IRS-assisted ISAC system in \cite{ref:IRS-ISAC-wavebeam1} \textcolor{black}{and} \cite{ref:IRS-ISAC-wavebeam2}, considering the \textcolor{black}{trade-off} between S\&C performance.
Note that the enhancement of S\&C performance builds on the accurate CSI in the above designs.
However, to the best of the authors' knowledge, the channel estimation problem in such IRS-assisted ISAC systems has not been investigated yet in the state-of-art literature.

This paper proposes a DL-based channel estimation approach \textcolor{black}{for the} IRS-assisted ISAC system.
The key contribution of this work is three-fold:
\begin{enumerate}[]
  \item A DL estimation framework, which involves two different {deep neural network (DNN)} architectures, is developed to estimate the S\&C channels. One \textcolor{black}{DNN} is employed at the ISAC base station (BS) to estimate the sensing channel between the ISAC BS and the target, while the other \textcolor{black}{one} is {assigned} to each downlink user equipment (UE) for the BS-IRS-UE channel estimation.
  \item The {generation of the} input-output pairs {for} the DNNs is designed. Moreover, the training \textcolor{black}{data} is enriched \textcolor{black}{through} augmentation to enhance the estimation performance for both S\&C channels.
  \item Numerical results demonstrate the substantial \textcolor{black}{improvements} achieved by the proposed approach over the benchmark \textcolor{black}{scheme} under various signal-to-noise ratio (SNR) conditions and \textcolor{black}{system parameters}.
\end{enumerate}

The rest of the paper is organized as follows: In Section \ref{sec:system}, the S\&C models of the IRS-assisted ISAC system are summarized. Section \ref{sec:DNN-CE-appraoch} introduces the proposed DL-based estimation approach. The simulation results and conclusions are provided in Sections \ref{sec:Simulation} and \ref{sec:Conclusion}\textcolor{black}{, respectively}.

\section{System Model}\label{sec:system}
\begin{figure}
\centering
\includegraphics[width=2.3in]{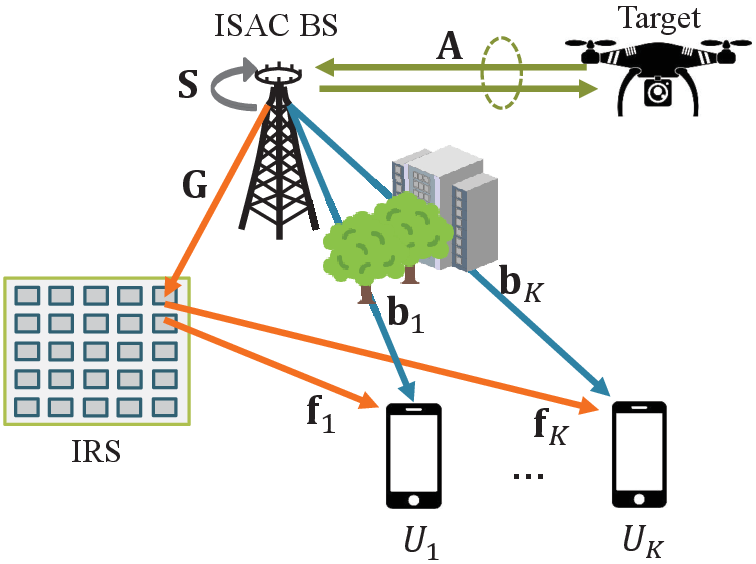}
\caption{IRS-assisted ISAC system model.$\qquad$$\qquad$$\qquad$$\qquad$$\qquad$$\qquad$$\qquad$$\quad$}
\label{fig:System}
\end{figure}

\textcolor{black}{Consider} an IRS-assisted ISAC system with \textcolor{black}{an} ISAC BS, IRS, \textcolor{black}{a} target, and $K$ downlink UEs, as illustrated in Fig. \ref{fig:System}, where $U_k$, $k\in\calN_1^K$, denotes the $k$-th UE.
Henceforth, $\calN_a^b=\{a,a+1,...b\}$ \textcolor{black}{represents the index set} from integer $a$ to $b$, and $a<b$.
The ISAC BS is equipped with $M$ transmit and $M$ \textcolor{black}{receive} antennas to sense the target by utilizing the echo signal from the BS-target-BS channel, $\bfA\in\bbC^{M\times M}$.
\textcolor{black}{With} the assistance of IRS consisting of $L$ passive reflecting elements, the ISAC BS communicates with the single-antenna downlink UEs through the direct and reflected channels.
\textcolor{black}{Let} $\bfb_k\in\bbC^{M\times 1}$, $\bfG\in\bbC^{M\times L}$, and $\bff_k\in\bbC^{L\times 1}$ denote the channel coefficients of the BS-$U_k$, BS-IRS, and IRS-$U_k$ links, respectively.
\textcolor{black}{Since} the ISAC BS transmits and receives \textcolor{black}{signals} simultaneously (i.e., full-duplex mode), the self-interference (SI) is induced to the ISAC BS from the SI channel, $\bfS\in\bbC^{M\times M}$.

\begin{figure}
\centering
\includegraphics[width=4in]{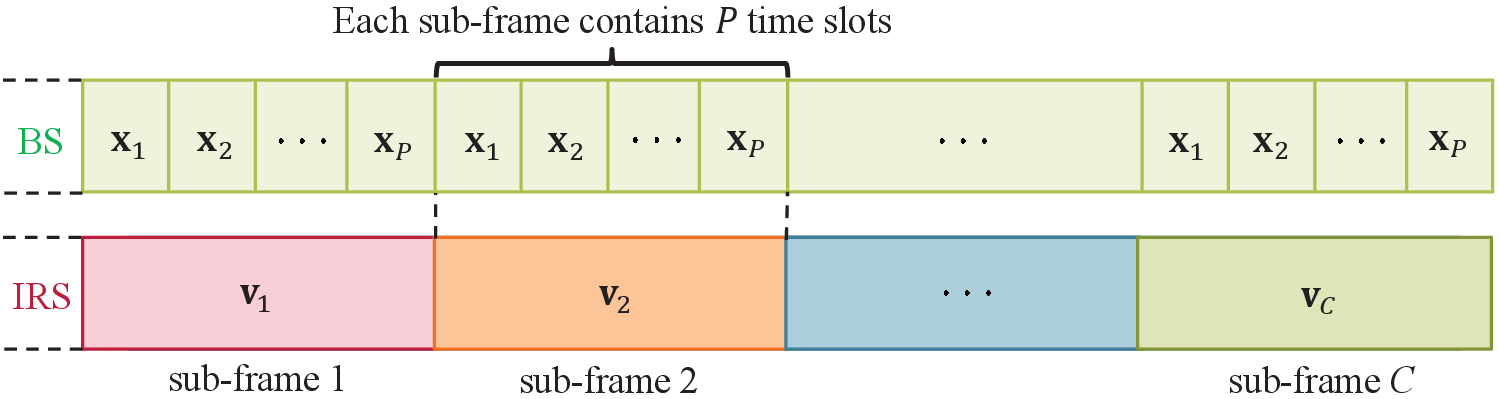}
\caption{Pilot transmission protocol.$\qquad$$\qquad$$\qquad$$\qquad$$\qquad$$\qquad$$\qquad$$\qquad$$\qquad$}
\label{fig:Pilot}
\end{figure}

\textcolor{black}{Fig. \ref{fig:Pilot} shows the} pilot transmission protocol \textcolor{black}{that} is designed to estimate the channels \textcolor{black}{of the IRS-assisted ISAC system}.
As seen \textcolor{black}{from} Fig. \ref{fig:Pilot}, the ISAC BS simultaneously transmits the pilot sequences to the target and UEs \textcolor{black}{in} $C$, $C\geq L$, sub-frames, and each sub-frame contains $P$, $P\geq M$, time slots.
The pilot signal matrix adopted in the $c$-th, $c\in\calN_1^C$, sub-frame is defined as $\bfX=[\bfx_1,\bfx_2,\ldots,\bfx_P]\in\bbC^{M\times P}$, where $\bfx_p\in\bbC^{M\times 1}$ is the pilot signal vector at the $p$-th time slot\textcolor{black}{, $p\in\calN_1^P$}.
Considering the demand for low pilot overhead, let $P=M$ and design $\bfX$ as \textcolor{black}{an} $M\times M$ discrete Fourier transform (DFT) matrix, where the $(n,q)$-th entry in $\bfX$ is expressed as ${\bf X}^{(n,q)}=\frac 1 {\sqrt{M}} e^{\jmath\frac{2\pi}{M}nq}$.
Furthermore, the \textcolor{black}{IRS} phase-shift vector in the $c$-th sub-frame is denoted by $\bfv_c\in\bbC^{L\times 1}$.
Referring to \cite{ref:DL-IRS-ChE-TWC}, the phase-shift matrix $\bfV=[\bfv_1,\bfv_2,\ldots,\bfv_C]\in\bbC^{L\times C}$ \textcolor{black}{is} a DFT matrix with $C=L$.
This has been proved to be an optimal choice of $\bfV$ to enhance the received signal power at UEs and ensure channel estimation accuracy \cite{ref:DL-IRS-ChE-TWC}.
\textcolor{black}{Accordingly}, it is noted that the ISAC BS sends the identical pilot sequences (i.e., $\bfX$), in each sub-frame.
\textcolor{black}{Thus}, \textcolor{black}{the IRS} phase-shift vector (i.e., $\bfv_c$) \textcolor{black}{is kept} constant within one sub-frame, whereas it varies among \textcolor{black}{the rest of the} sub-frames.

\subsection{Sensing Model}
Based on the pilot transmission protocol in Fig. \ref{fig:Pilot}, the received sensing signal at the $p$-th time slot in the $c$-th sub-frame\textcolor{black}{, $\bfy_{c,p}$,} at the ISAC BS is given by
\begin{align}
& \bfy_{c,p} = \underbrace{ \bfA^\rmH\bfx_{p} }_{\text {Sensing signal}} + \underbrace{ \bfS^\rmH\bfx_{p} }_{\text {Residual SI}} + \bfn_{c,p}, 
\label{eq:y_cp}
\end{align}
where $\bfn_{c,p}\sim{\cal CN}(0,\sigma^2\bfI_M)$ denotes the noise at the ISAC BS that follows \textcolor{black}{the} zero-mean complex Gaussian distribution {of} variance $\sigma^2$\textcolor{black}{,} {with} $\bfI_M$ \textcolor{black}{as} an identity {matrix} of size $M$.
Refer to the radar channel model in \cite{ref:ChModel-b,ref:IRS-ISAC-wavebeam2}, the sensing channel is modeled as
\begin{align}
\bfA = \alpha_\rmS \bfa(\theta_\rmS)\bfa(\theta_\rmS)^\rmT, 
\label{eq:A_model}
\end{align}
where $\alpha_\rmS$ represents the complex-valued reflection coefficient of the target.
\textcolor{black}{Here,} $\bfa(\theta_\rmS)$ \textcolor{black}{is} the steering vector of the BS transmit antenna array associated with the target's azimuth angle $\theta_\rmS$ \textcolor{black}{that} can be expressed as
\begin{align}
\bfa(\theta_\rmS) = [1, e^{\jmath\frac{2\pi {d_\rmB}}{\lambda}\sin(\theta_\rmS)}, \ldots, e^{\jmath\frac{2\pi {d_\rmB}}{\lambda}({M}-1)\sin(\theta_\rmS)}]^\rmT,
\label{eq:a_thetaj}
\end{align}
where $d_\rmB$ and $\lambda$ are the antenna spacing of ISAC BS and the signal wavelength, respectively.
Without loss of generality, \textcolor{black}{assume} \textcolor{black}{that} $d_\rmB=\frac \lambda 2$.
\textcolor{black}{The} propagation environment between the transmit and received antennas of the ISAC BS is assumed to be slow-changing \cite{ref:FD-BS-slowchanging}.
Hence, the SI channel, $\bfS$, can be pre-estimated at the ISAC BS, and the residual SI in \eqref{eq:y_cp} is \textcolor{black}{compensated} before estimating $\bfA$.

\subsection{Communication Model}
For the downlink $U_k$, the received signal at the $p$-th time slot in the $c$-th sub-frame\textcolor{black}{, $z_{k,c,p}$,} can be expressed as
\begin{align}
& z_{k,c,p} = \underbrace{ \Big( \bfb_k^\rmH + \bff_k^\rmH \rmdiag\{\bfv_c^\rmH\} \bfG^\rmH \Big)\bfx_{p} }_{\text {Downlink communication signal}} + w_{k,c,p}, 
\label{eq:r_kcp}
\end{align}
where $\bfv_c=[\beta_c e^{\jmath\varphi_{c,1}}, \beta_c e^{\jmath\varphi_{c,2}}, \ldots, $ $\beta_c e^{\jmath\varphi_{c,L}}]^\rmT$ with $\beta_c\in[0,1]$\textcolor{black}{,} and $\varphi_{c,\ell}\in[0,2\pi)$ \textcolor{black}{with} $\ell\in\calN_1^L$ \textcolor{black}{are} the amplitude and phase-shift of the $\ell$-th IRS element, respectively.
The noise $w_{k,c,p}$ follows $\mathcal{CN}(0,\varsigma^2)$ with \textcolor{black}{zero-mean and} variance $\varsigma^2$.
From \eqref{eq:r_kcp}, it is obvious that $\bfG\rmdiag\{\bfv_c\}\bff_k=\bfG\rmdiag\{\bff_k\}\bfv_c$, {where $\rmdiag\{\cdot\}$ converts a vector into a diagonal matrix with diagonal elements that are the same as the original vector elements.} \textcolor{black}{Thus}, the equivalent reflected channel of the BS-IRS-$U_k$ link can be expressed as $\bfB_k = \bfG\rmdiag\{\bff_k\}\in\bbC^{M\times L}$.
\textcolor{black}{The} direct BS-$U_k$ channel, $\bfb_k$, tends to be blocked by obstacles (e.g., buildings \textcolor{black}{or} trees), \textcolor{black}{and} the reflected channel is considered \cite{ref:ChModel-refpower}. 
As such, \eqref{eq:r_kcp} can be rewritten as
\begin{align}
& z_{k,c,p} = \underbrace{ \bfv_c^\rmH\bfB_k^\rmH \bfx_{p} }_{\text {Downlink communication signal}} + w_{k,c,p}. 
\label{eq:r_kcp_B}
\end{align}
To model $\bfB_k$, the channels of BS-IRS and IRS-$U_k$ links, $\bfG$ and $\bff_k$, are modeled as Rician fading \cite{ref:DL-IRS-ChE-TWC,ref:ChModel-refpower}.
\textcolor{black}{Therefore}, $\bfG$ is \textcolor{black}{given} by
\begin{align}
\bfG = \sqrt{\frac {K_\rmBI}{K_\rmBI+1}} \bfG_{\rmLoS}
+ \sqrt{\frac {1}{K_\rmBI+1}} \bfG_{\rmNLoS},
\label{eq:H_rician}
\end{align}
where $K_\rmBI$ represents the Rician factor.
{Here,} $\bfG_{\rmLoS}$ and $\bfG_{\rmNLoS}$ are the line-of-sight {(LoS)} component and {non-LoS (NLoS)} component, respectively.
In addition, define $\bfG_{\rmLoS}=\bfa(\theta_\rmB)\bfa(\theta_\rmI)^\rmH$, where the steering vectors $\bfa(\theta_\rmB)$ and $\bfa(\theta_\rmI)$ are formulated {similarly as in} \eqref{eq:a_thetaj}.
{Particularly}, $\bfa(\theta_\rmB)$ is associated with the angle of departure (AoD) from the BS to the IRS, $\theta_\rmB$, and the parameters of the BS antenna array (i.e., $M$ and $d_\rmB$){,} while $\bfa(\theta_\rmI)$ is related to the angle of arrival (AoA) at the IRS, $\theta_\rmI$, as well as the parameters of the IRS (i.e., $L$ and inter-element spacing $d_\rmI=\frac \lambda 2$).
The IRS-$U_k$ channel, $\bff_k$, can be modeled similarly as in \eqref{eq:H_rician}.

In the following, we aim to estimate the S\&C channels, $\{\bfA, \bfB_k\}$, $k\in\calN_1^K$, \textcolor{black}{for the} IRS-assisted ISAC system based on the designed pilot transmission protocol.

\section{\textcolor{black}{Proposed} DL-based Estimation Approach}\label{sec:DNN-CE-appraoch}
This section proposes a practical channel estimation approach based on DL.
The generation of the input-output pairs for the DL networks is firstly designed.
On this basis, a DNN-based estimation framework is then developed.

\subsection{\textcolor{black}{Input-Output Pairs} Design}
\label{sec:IO-design}
\subsubsection{Design for Sensing Channel}
To generate the input-output pairs of the \textcolor{black}{DL} for the sensing channel estimation, $P$ received signal vectors in the $c$-th sub-frame at the ISAC BS (i.e., $\bfy_{c,p}$, $p\in\calN_1^P$) are firstly stacked.
Since the residual SI is assumed to be compensated, the matrix form of \eqref{eq:y_cp} is
\begin{align}
\bfY_c = \bfA^\rmH\bfX + \bfN_c, \quad c\in\calN_{1}^{C},
\label{eq:Y_c}
\end{align}
where $\bfY_c=[\bfy_{c,1},\bfy_{c,2},\ldots,\bfy_{c,P}]\in\bbC^{M\times P}$ and $\bfN_c=[\bfn_{c,1},\bfn_{c,2},\ldots,\bfn_{c,P}]\in\bbC^{M\times P}$.
\textcolor{black}{Therefore}, the input of the \textcolor{black}{DL}, $\bfR^\rmS$, is generated by utilizing the received sensing signals in \eqref{eq:Y_c} as
\begin{align}
& \bfR^\rmS = \Big[\Re\{{\rm{vec}} [\bfY_1,\bfY_2,\ldots,\bfY_C] \},  \Im\{{\rm{vec}} [\bfY_1,\bfY_2,\ldots,\bfY_C] \}\Big]^\rmT,
\end{align}
where $\Re\{\cdot\}$ and $\Im\{\cdot\}$ represent the operations of extracting the real and imaginary parts of a complex vector, respectively.
$\rm{vec}[\cdot]$ denotes the operation of converting a matrix to a vector.
Correspondingly, the output of the \textcolor{black}{DL}, $\bfH^\rmS$, is constructed by the ground truth of the sensing channel $\bfA$ \textcolor{black}{as}
\begin{align}
\bfH^{\rmS} = \Big[\Re\{{\rm{vec}}[\bfA]\}, \Im\{{\rm{vec}}[\bfA]\}\Big]^\rmT.
\end{align}

Based on the model in \eqref{eq:Y_c}, the least-squares (LS) estimator is introduced as a benchmark.
Define the LS estimation result of the sensing channel as $\bar\bfA^{\rm LS}$, which is obtained by
\begin{align}
\bar\bfA^{\rm LS} = \mathbb E\big\{ ( \bfY_c\bfX^\dag)^\rmH\big\} = \bfA + \mathbb E\big\{ \bar\bfN_c^\rmH\big\},
\end{align}
where $\bfX^\dag=\bfX^\rmH(\bfX\bfX^\rmH)^{-1}$ denotes the pseudoinverse of $\bfX$\textcolor{black}{,} $\bar\bfN_c=\bfN_c\bfX^\dag$, \textcolor{black}{and $\mathbb E\{\cdot\}$ is the expectation operation}.

\subsubsection{Design for Communication Channel}
To generate the input-output \textcolor{black}{pairs} of the \textcolor{black}{DL} for the communication channel estimation, the vector form of \eqref{eq:r_kcp_B} is firstly given by
\begin{align}
\bfz_{k,c} = \bfv_c^\rmH\bfB_k^\rmH \bfX + \bfw_{k,c}, \quad k\in\calN_1^K, \quad c\in\calN_{1}^{C},
\label{eq:r_kc}
\end{align}
where $\bfz_{k,c}=[z_{k,c,1},z_{k,c,2},\ldots,z_{k,c,P}]\in\bbC^{1\times P}$ and $\bfw_{k,c}=[w_{k,c,1},w_{k,c,2},\ldots,w_{k,c,P}]\in\bbC^{1\times P}$.
Then, by utilizing the received communication signal in \eqref{eq:r_kc}, the input of the \textcolor{black}{DL} at the downlink $U_k$, $\bfR^{U_k}$, is generated as
\begin{align}
& \bfR^{U_k} = \Big[\Re\Big\{\big[\bfz_{k,1},\bfz_{k,2},\ldots\bfz_{k,C}\big]\Big\}, \Im\Big\{\big[\bfz_{k,1},\bfz_{k,2},\ldots\bfz_{k,C}\big]\Big\}\Big]^{\rmT}.
\end{align}
The corresponding output of the \textcolor{black}{DL} at the downlink $U_k$, $\bfH^{U_k}$, is generated by the ground truth of the communication channel $\bfB_k$ \textcolor{black}{as}
\begin{align}
\bfH^{U_k} = \Big[\Re\{{\rm{vec}}[\bfB_k]\}, \Im\{{\rm{vec}}[\bfB_k]\}\Big]^\rmT.
\end{align}

According to the model in \eqref{eq:r_kc}, the LS estimator is adopted as a benchmark at the downlink $U_k$, $k\in\calN_1^K$.
By separating the orthogonal pilot matrix $\bfX$ from $\bfz_{k,c}$ in \eqref{eq:r_kc}, the derived $\tilde\bfz_{k,c}$ is formulated as
\begin{align}
\tilde\bfz_{k,c}
& = \bfz_{k,c} \bfX^\dag \notag\\
& = \bfv_c^\rmH\bfB_k^\rmH + \tilde\bfw_{k,c}, \quad k\in\calN_1^K, \quad c\in\calN_{1}^{C}, 
\label{eq:r_bar_S2}
\end{align}
where $\tilde\bfw_{k,c} = \bfw_{k,c}\bfX^\dag$.
Then, \textcolor{black}{from} the first to the $C$-th sub-frames, the matrix form of \eqref{eq:r_bar_S2} is written as
\begin{align}
\tilde\bfZ_k = \bfV^\rmH \bfB_k^\rmH + \tilde\bfW_k, \quad k\in\calN_1^K, 
\end{align}
where $\tilde\bfZ_k=[\tilde\bfz_{k,1}^\rmT,\tilde\bfz_{k,2}^\rmT,\ldots,\tilde\bfz_{k,C}^\rmT]^\rmT\in\bbC^{C\times M}$ and $\tilde\bfW_k=[\tilde\bfw_{k,1}^\rmT,\tilde\bfw_{k,2}^\rmT,\ldots,\tilde\bfw_{k,C}^\rmT]^\rmT\in\bbC^{C\times M}$.
Hence, the LS estimate of the BS-IRS-$U_k$ channel is
\begin{align}
\bar\bfB_k^{\rm LS} = \tilde\bfZ_k^\rmH \bfV^\dag, \quad k\in\calN_1^K.
\end{align}

\subsubsection{Training \textcolor{black}{Dataset} Generation}
The training \textcolor{black}{dataset of} the \textcolor{black}{DL} is constructed by the designed input-output pairs.
For the sensing channel estimation, define the training \textcolor{black}{dataset} as
\begin{align}
& (\calR^\rmS,\calH^\rmS)=\Big\{ \big(\bfR^\rmS_{(1,1)},\bfH^\rmS_{(1)}\big), \big(\bfR^\rmS_{(1,2)},\bfH^\rmS_{(1)}\big),\ldots, \notag\\
& \big(\bfR^\rmS_{(1,U)},\bfH^\rmS_{(1)}\big),
\big(\bfR^\rmS_{(2,1)},\bfH^\rmS_{(2)}\big), \ldots, \big(\bfR^\rmS_{(V,U)},\bfH^\rmS_{(V)}\big)  \Big\},
\label{eq:IO_set}
\end{align}
where $\big(\bfR^\rmS_{(v,u)},\bfH^\rmS_{(v)}\big)$ denotes the $(v,u)$-th, $v\in\calN_1^V$, $u\in\calN_1^U$, training sample.
Based on the data augmentation, the training samples in this {dataset} are generated by adopting $V$ received original signals (i.e., sensing signal in \eqref{eq:Y_c}), and $U-1$ copies of the $v$-th, $v\in\calN_1^V$, are generated by introducing synthetic additive white Gaussian noise to the original channel $\bfA$ with $\text{SNR}_\text{ch} =\frac{{\cal P}_\text{ch} }{\sigma_\text{ch}^2}$.
\textcolor{black}{Here,} ${\cal P}_\text{ch}$ and $\sigma_\text{ch}^2$ represent the power of the original channel $\bfA$ and the synthetic noise, respectively \cite{ref:DL-IRS-ChE-WCL}.
As such, these noise corrupted channels can be adopted to obtain the copy signals and then generate $U-1$ copy samples, (i.e., $\big(\bfR^\rmS_{(v,u)},\bfH^\rmS_{(v)}\big)$, $v\in\calN_1^V$, $u\in\calN_2^U$).
\textcolor{black}{In addition to} their original versions (i.e., $\big(\bfR^\rmS_{(v,1)},\bfH^\rmS_{(v)}\big)$, $v\in\calN_1^V$), the \textcolor{black}{dataset} in \eqref{eq:IO_set} is constructed.
It is worth mentioning that these copy samples are generated to enrich the training \textcolor{black}{dataset to} enhance the estimation performance of the DL network.
Moreover, the training \textcolor{black}{dataset} for the communication channel estimation, $(\calR^{U_k},\calH^{U_k})$, $k\in\calN_1^K$, \textcolor{black}{is} generated similarly as in \eqref{eq:IO_set} at the downlink UE.

\subsection{Proposed DL Estimation Framework}
\begin{figure}
\centering
\subfigure[ ]
{\begin{minipage}[b]{0.7\textwidth}
\includegraphics[width=1\textwidth]{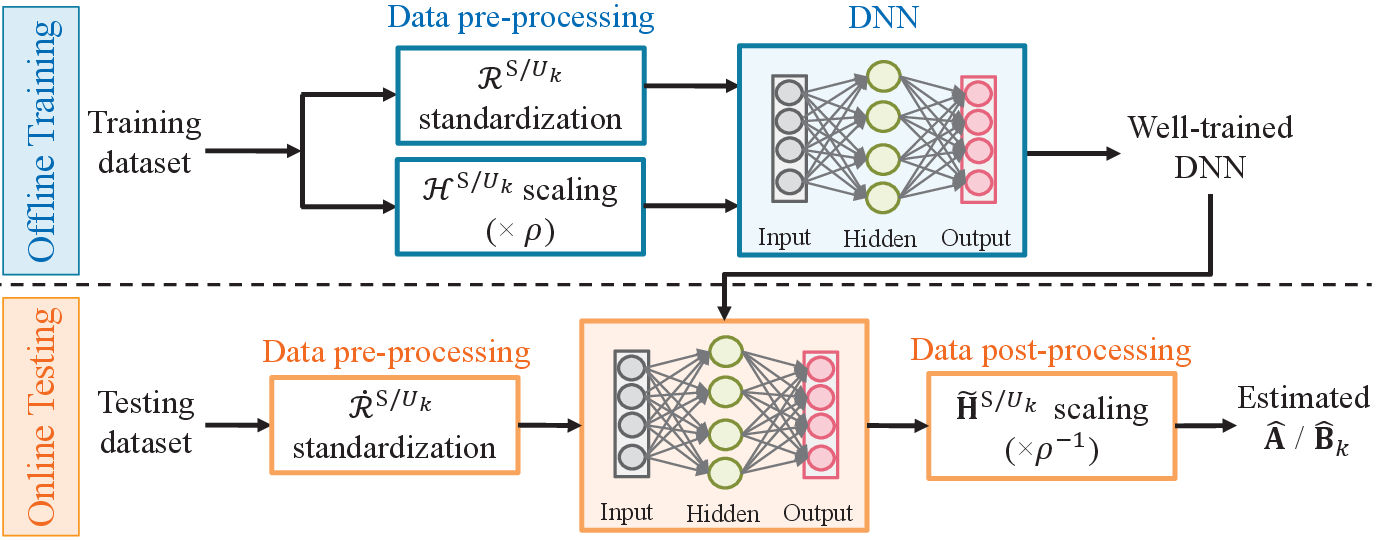}
\end{minipage}}

\subfigure[ ]
{\begin{minipage}[b]{0.6\textwidth}
\includegraphics[width=1\textwidth]{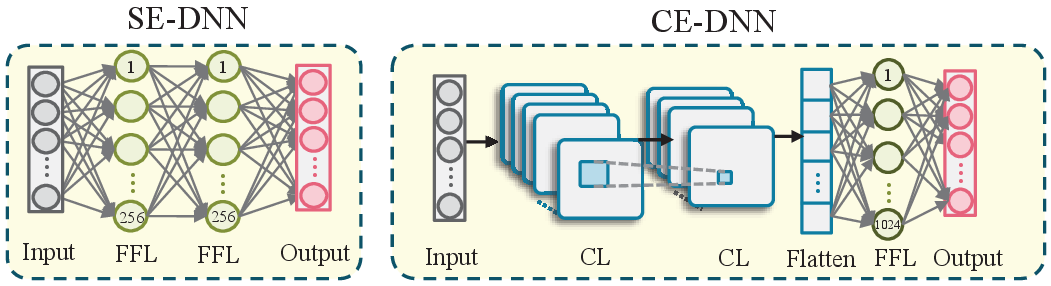}
\end{minipage}}
\caption{The proposed DL estimation framework: (a) Offline training and online testing phases, (b) \textcolor{black}{Architectures} of the SE-DNN and CE-DNN.}
\label{fig:DNN_Diag}
\end{figure}

Based on the designed input-output pairs, \textcolor{black}{the} DNN-based estimation framework that consists of the offline training and online testing phases is developed in Fig. \ref{fig:DNN_Diag}.

\subsubsection{Offline Training}
For both S\&C channel estimation, the offline training phase is composed of the data pre-processing and the network training procedures, as depicted in Fig. \ref{fig:DNN_Diag}(a).
The data pre-processing is firstly performed on the training {dataset}, including the input standardization and the output scaling with {a} factor \textcolor{black}{of} \textcolor{black}{$\rho=10^4$}.
Then, the networks in the developed framework are trained by the pre-processed samples to obtain the well-trained networks.
Two different DNN architectures are designed to {form} this framework.
{The first} \textcolor{black}{DNN} is devised for the {sensing} channel {estimation}, namely {sensing estimation DNN (SE-DNN)}, while the other \textcolor{black}{one} is employed for the {communication} channel {estimation}, referred to as \textcolor{black}{the} {communication estimation DNN (CE-DNN)}.

Here, the detailed offline training \textcolor{black}{procedure} for the sensing channel estimation is \textcolor{black}{explained}.
Given the \textcolor{black}{dataset} $(\calR^\rmS,\calH^\rmS)$ in \eqref{eq:IO_set}, the $(v,u)$-th pre-processed input-output \textcolor{black}{pairs} in this \textcolor{black}{dataset} is obtained as $\big(\bar{\bfR}^\rmS_{(v,u)},\bar{\bfH}^\rmS_{(v)}\big)$.
Then, in the \textcolor{black}{DNN} training \textcolor{black}{phase}, the designed SE-DNN approximates $\bar{\bfH}^\rmS_{(v)}$ as
\begin{align}
\bar{\bfH}^\rmS_{(v)} \approx f^\rmS \Big(\bar{\bfR}^\rmS_{(v,u)}; \Theta^\rmS \Big), \quad v\in\calN_1^V, \quad u\in\calN_1^U,
\end{align}
where $f^\rmS(\cdot; \Theta^\rmS)$ denotes the SE-DNN \textcolor{black}{function} and $\Theta^\rmS$ represents all its \textcolor{black}{hyperparameters. The SE-DNN updates $\Theta^\rmS$ by minimizing} the mean square error (MSE) \textcolor{black}{of} the loss function, \textcolor{black}{${\cal F}^\rmS$, as}
\begin{align}
{\cal F}^\rmS = \frac 1 {VU} \sum_{v\in\calN_1^{V}, u\in\calN_1^{U}} \bigg(f^\rmS \Big(\bar{\bfR}^\rmS_{(v,u)}; \Theta^\rmS\Big)-\bar{\bfH}_v^\rmS\bigg)^2.
\label{eq:loss}
\end{align}
The offline training of the CE-DNN performs similarly to the SE-DNN.

Fig. \ref{fig:DNN_Diag}(b) \textcolor{black}{illustrates} the architectures of the SE-DNN and CE-DNN.
Both consist of the input, hidden, and output layers.
\textcolor{black}{Particularly}, the hidden layers in the SE-DNN are \textcolor{black}{formed from} two feed-forward layers (FFLs) with {\textit {tanh}} activation functions.
Since the propagation environment of the reflected communication channel is more complicated than that of the sensing channel, the CE-DNN contains three hidden layers \textcolor{black}{that consist of} two convolutional layers (CLs) \textcolor{black}{followed by} a flatten layer \textcolor{black}{and FFL at the end}.
The CLs utilize {\textit {tanh}} activation functions, while the FFL adopts \textcolor{black}{a} {\textit {linear}} one.
The advantages of this design are attributed to two aspects\textcolor{black}{, as follows}.
\textcolor{black}{First}, the neurons in the CL are connected locally, and partial connections among them share the same weight.
Hence, the CL effectively reduces the network parameters and accelerates the convergence compared to the fully connected network \cite{ref:CNN-merit}.
\textcolor{black}{Second}, the FFL is further combined with the CL, promoting the channel estimation performance and the generalization capacity of the CE-DNN.
For both SE-DNN and CE-DNN, \textcolor{black}{the} {\textit{Adam}} optimizer is adopted to update the \textcolor{black}{DNN} parameters with \textcolor{black}{a} learning rate of \textcolor{black}{$2\times 10^{-4}$} and minibatch transitions of size $200$.
Moreover, a stopping criterion is applied in the training process, in which the training \textcolor{black}{stops} if the validation loss does not improve in $5$ consecutive epochs, or the number of the training epochs \textcolor{black}{reaches} $300$.
The detailed hyperparameters of the SE-DNN and CE-DNN are summarized in Table \ref{table:DNN}.

\begin{table}[t] 
\centering
\caption{Hyperparameters of SE-DNN and CE-DNN} \label{table:DNN}
\begin{tabular}{p{1cm}<{\centering}| p{2cm}<{\centering}| p{2cm}<{\centering}| p{2cm}<{\centering}| p{3cm}<{\centering}}
\hline
                            & {\bf Layer type} & {\bf Tensor size} & {\bf Kernel size} & {\bf Activation function} \\
\hline
\hline
\multirow{4}{*}{\bf SE-DNN} & Input            & $2MPC$            & -             & -    \\ \cline{2-5}
                            & FFL              & $256$             & -             & \textcolor{black}{\textit{tanh}} \\ \cline{2-5}
                            & FFL              & $256$             & -             & \textcolor{black}{\textit{tanh}} \\ \cline{2-5}
                            & Output           & $2M^2$            & -             & -    \\ \cline{1-5}
\multirow{5}{*}{\bf CE-DNN} & Input            & $2PC$             & -             & -    \\ \cline{2-5}
                            & CL               & $128$             & $4\times 1$   & \textcolor{black}{\textit{tanh}} \\ \cline{2-5}
                            & CL               & $64$              & $4\times 1$   & \textcolor{black}{\textit{tanh}} \\ \cline{2-5}
                            & FFL              & $1024$            & -             & \textcolor{black}{\textit{linear}} \\ \cline{2-5}
                            & Output           & $2ML$             & -             & -     \\
\hline
\end{tabular}
\end{table} 



\subsubsection{Online Testing}
Corresponding to the offline training, the online testing phase for the sensing channel estimation is illustrated {in} Fig. \ref{fig:DNN_Diag}(a).
{The} testing dataset, $\dot{\calR}^\rmS$, is pre-processed to obtain standardized samples \textcolor{black}{that are} denoted by $\tilde{\bfR}^\rmS$.
Then, the output of the SE-DNN, $\tilde{\bfH}^\rmS$, can be expressed as
\begin{align}
\tilde{\bfH}^\rmS = f^\rmS(\tilde{\bfR}^\rmS;\hat\Theta^\rmS),
\end{align}
where $\hat\Theta^\rmS$ represents the \textcolor{black}{hyperparameters} of the \textcolor{black}{trained} SE-DNN.
After that, the sensing channel is estimated as $\hat\bfA$ by scaling $\tilde{\bfH}^\rmS$ with \textcolor{black}{the} factor \textcolor{black}{$\rho^{-1}$} and performing simple matrix operations.
Similarly, with the testing {dataset, $\dot{\calR}^{U_k}$,} at the downlink $U_k$, \textcolor{black}{the} communication channel is estimated as $\hat\bfB_k$ by the \textcolor{black}{trained} CE-DNN, $f^{U_k}(\cdot;\hat\Theta^{U_k})$.
The proposed DNN-based channel estimation algorithm is summarized in Algorithm \ref{algo:DNN-CE}, where $t^\rmS$ and $t^{U_k}$ denote the indices of the training epoch used in the \textcolor{black}{S\&C} channel estimation, respectively.

\newlength\myindent 
\setlength\myindent{2em}
\newcommand\bindent{
    \begingroup
    \setlength{\itemindent}{\myindent}
    \addtolength{\algorithmicindent}{\myindent} }
\newcommand\eindent{\endgroup}

\begin{algorithm}[!t]
\caption{DNN-based \textcolor{black}{Channel Estimation Algorithm}.}
{{\bf Initialize:} $t^\rmS=0$ and $t^{U_k}=0$, $k\in\calN_1^K$; } \\
{\bf Offline training:}

\begin{algorithmic}[1]  
    \STATE {\bf Generate} $(\calR^\rmS,\calH^\rmS)$ and $(\calR^{U_k},\calH^{U_k})$ according to Section \ref{sec:IO-design};
    \STATE {\bf Pre-process} $(\calR^\rmS,\calH^\rmS)$ and $(\calR^{U_k},\calH^{U_k})$;
    \STATE {\bf Input} the pre-processed samples, $\big(\bar{\bfR}^\rmS_{(v,u)},\bar{\bfH}^\rmS_{(v)}\big)$ and $\big(\bar{\bfR}^{U_k}_{(v,u)},\bar{\bfH}^{U_k}_{(v)}\big)$, to the proposed SE-DNN and CE-DNN, respectively; 

    \WHILE {the number of epochs does not reach $300$ or the validation accuracy improves in $5$ consecutive epochs}
      \STATE {\bf Update} $\Theta^\rmS$ by minimizing ${\cal F}^\rmS$ at the ISAC BS and $\Theta^{U_k}$ by minimizing ${\cal F}^{U_k}$ at the downlink $U_k$, using \textcolor{black}{the} {\textit{Adam}} optimizer;
      \STATE {\bf Set} $t^\rmS \leftarrow t^\rmS+1$ and $t^{U_k} \leftarrow t^{U_k}+1$;
    \ENDWHILE
    \STATE {\bf Output} the \textcolor{black}{trained} SE-DNN, $f^\rmS(\cdot \,; \hat\Theta^\rmS)$, at the ISAC BS and the \textcolor{black}{trained} CE-DNN, $f^{U_k}(\cdot \,; \hat\Theta^{U_k})$, at the downlink $U_k$.

\end{algorithmic}

{\bf Online testing:}
\begin{algorithmic}[1]  

  \STATE {\bf Input:} Testing \textcolor{black}{dataset $\dot{\calR}^\rmS$ and $\dot{\calR}^{U_k}$}, $k\in\calN_1^K$;
  \begin{ALC@g}
    \STATE {\bf Estimate} $\hat\bfA$ by using $f^\rmS(\cdot \,; \hat\Theta^\rmS)$ with the \textcolor{black}{standardized sample $\tilde{\bfR}^\rmS$} at the ISAC BS;
    \STATE {\bf Estimate} $\hat\bfB_k$ by using $f^{U_k}(\cdot \,; \hat\Theta^{U_k})$ with the \textcolor{black}{standardized sample $\tilde{\bfR}^{U_k}$} at the downlink $U_k$;
  \end{ALC@g}
  \STATE {\bf Output:} Estimated channels $\{\hat\bfA, \hat\bfB_k\}$, $k\in\calN_1^K$.

\end{algorithmic}
\label{algo:DNN-CE}
\end{algorithm}


\section{Simulation Results}\label{sec:Simulation}
This section assesses the performance of the proposed DL-based channel estimation approach for the IRS-assisted ISAC system, considering the LS estimator as benchmark \textcolor{black}{for the comparison}.
In simulations, $K=3$, $M=4$, and $L=30$ unless further specified.
For the sensing channel $\bfA$, the {phase-shift} of the reflection {coefficients are} uniformly distributed {from} $[0,2\pi)$ and  $|\alpha_\rmS|=1$ \cite{ref:ChModel-b}.
The azimuth angle of the target is set to $\theta_\rmS=-\frac{2\pi}{3}$.
For the communication channels, the AoD and \textcolor{black}{AoA} associated with \textcolor{black}{$\bfG_\rmLoS$} are set to $\theta_\rmB=\theta_\rmI=\frac{\pi}{3}$.
The Rician factor of the BS-IRS link is $K_\rmBI=10$, while that of the IRS-$U_k$ link is $K_{\rmI U_k}=0$ \cite{ref:DL-IRS-ChE-TWC}.
Moreover, the path losses of the S\&C channels {are} $\zeta_\rmS = \zeta_0(\frac{d_\rmS}{d_0})^{-\gamma_\rmS}$, $\zeta_\rmBI = \zeta_0(\frac{d_\rmBI}{d_0})^{-\gamma_\rmBI}$, and $\zeta_{\rmI U_k} = \zeta_0(\frac{d_{\rmI U_k}}{d_0})^{-\gamma_{\rmI U_k}}$, where $\zeta_0=-30\,{\rm dBm}$ represents the path loss at the reference distance $d_0=1\,\rm m$.
The distances of BS-target-BS, BS-IRS, and IRS-$U_k$ are set to \textcolor{black}{$d_\rmS=140\,\rm m$, $d_\rmBI=50\,\rm m$, and $d_{\rmI U_k}=2\,\rm m$}, respectively.
The corresponding path loss exponents are $\gamma_\rmS=3$, $\gamma_\rmBI=2.3$, and $\gamma_{\rmI U_k}=2$, respectively \cite{ref:ChModel-refpower}.
\textcolor{black}{The} transmit power of the ISAC BS is set to $\mathcal{P}_0=20\,\rm dBm$ \cite{ref:ChModel-power}.

The hyperparameters of the \textcolor{black}{proposed} SE-DNN and CE-DNN have been summarized in Table \ref{table:DNN}.
For both S\&C channel estimation, the training \textcolor{black}{dataset} size is set to \textcolor{black}{$T_{\rm off}=VU=10^4$} for each SNR condition, with \textcolor{black}{$V=10^3$}, $U=10$, and \textcolor{black}{$\text{SNR}_\text{ch}=30\,\rm dB$}.
Adopt $90\%$ of the \textcolor{black}{dataset} for training while the rest for testing.
Furthermore, another \textcolor{black}{$T_{\rm on}=10^3$} data samples are tested under each SNR condition in the online testing phase.
The SNRs at the ISAC BS and the downlink $U_k$ are respectively defined as $\text{SNR}_\rmB = \frac{{\cal P}_\rmB}{\sigma^2}$ and $\text{SNR}_{U_k} = \frac{{\cal P}_{U_k}}{\varsigma^2}$, where \textcolor{black}{${\cal P}_\rmB=\mathcal{P}_0\zeta_\rmS$} and \textcolor{black}{${\cal P}_{U_k}=\mathcal{P}_0\zeta_\rmBI\zeta_{\rmI U_k}$} refer to their received signal power.
To evaluate the estimation performance, the normalized MSE (NMSE) is employed as performance metric (e.g., NMSE for sensing channel $\bfA$ is denoted by {$\text{NMSE}=\mathbb{E}\big\{ \frac{\|\hat \bfA - \bfA\|_F^2}{\| \bfA \|_F^2} \big\}$}, with $\|\cdot\|_F$ as the $F$-norm of a matrix).

\begin{figure}
\centering
\includegraphics[width=3.7in]{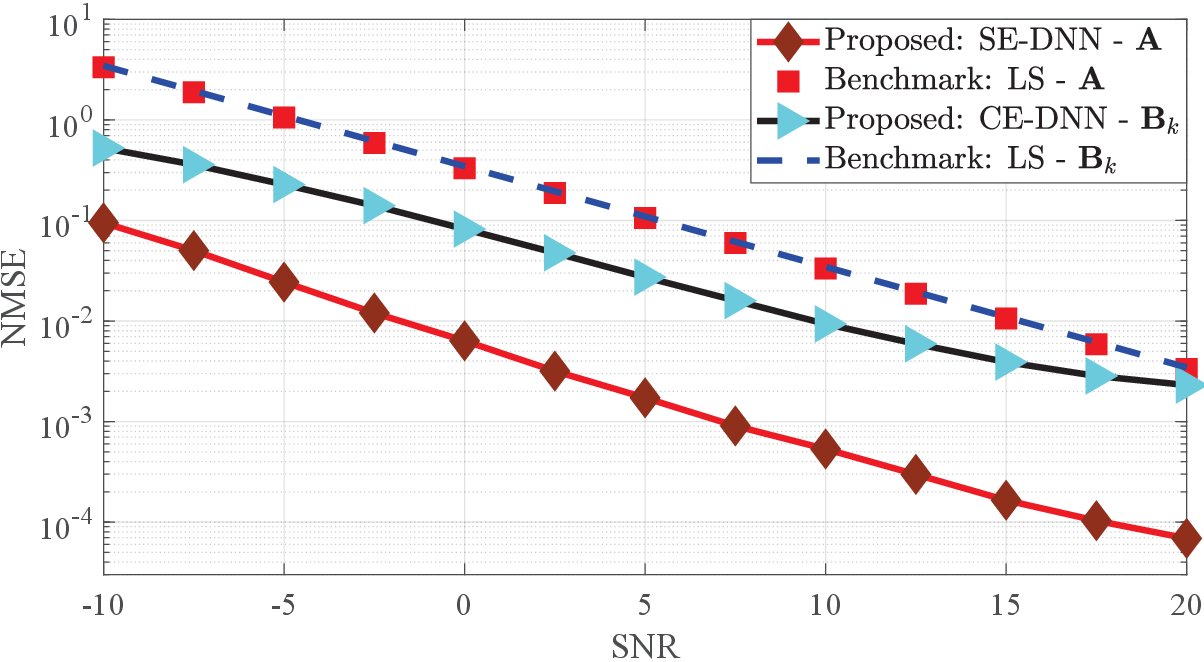}
\caption{NMSE of S\&C channel estimation versus SNR \textcolor{black}{for} $M=4$, $L=30$, and $K=3$.}
\label{fig:NMSE_SNR}
\end{figure}

Fig. \ref{fig:NMSE_SNR} studies the effect of the SNR on the S\&C channel estimation performance.
Let \textcolor{black}{$\text{SNR}= [10,20]\,\rm{dB}$} with a step size of $5\,\rm{dB}$ in the offline training phase, whereas \textcolor{black}{$\text{SNR} = [-10,20]\,\rm{dB}$} with a step size of $2.5\,\rm{dB}$ \textcolor{black}{for} online testing.
As can be observed, the proposed approach outperforms the benchmark \textcolor{black}{scheme} for both $\bfA$ and $\bfB_k$.
\textcolor{black}{Moreover, it is noted that compared to the benchmark scheme, the proposed approach obtains $15\,\rm dB$ SNR improvement at $\text{NMSE}=10^{-1}$ for the estimation of $\bfA$, while $5\,\rm dB$ SNR improvement at $\text{NMSE}=10^{-1}$ for the estimation of $\bfB_k$.}
The reason is that the sensing channel model in \eqref{eq:A_model} is simpler than the communication one, and thus, the mapping of the input-output pair $(\bfR^\rmS,\bfH^\rmS)$ is more straightforward to be learned by the \textcolor{black}{proposed} SE-DNN.
Apart from this, Fig. \ref{fig:NMSE_SNR} also reveals that the proposed approach possesses considerable generalization ability since the DNNs are trained under several SNR conditions \textcolor{black}{that} can be applied to a wide range of SNR regions.

\begin{figure}
\centering
\includegraphics[width=3.7in]{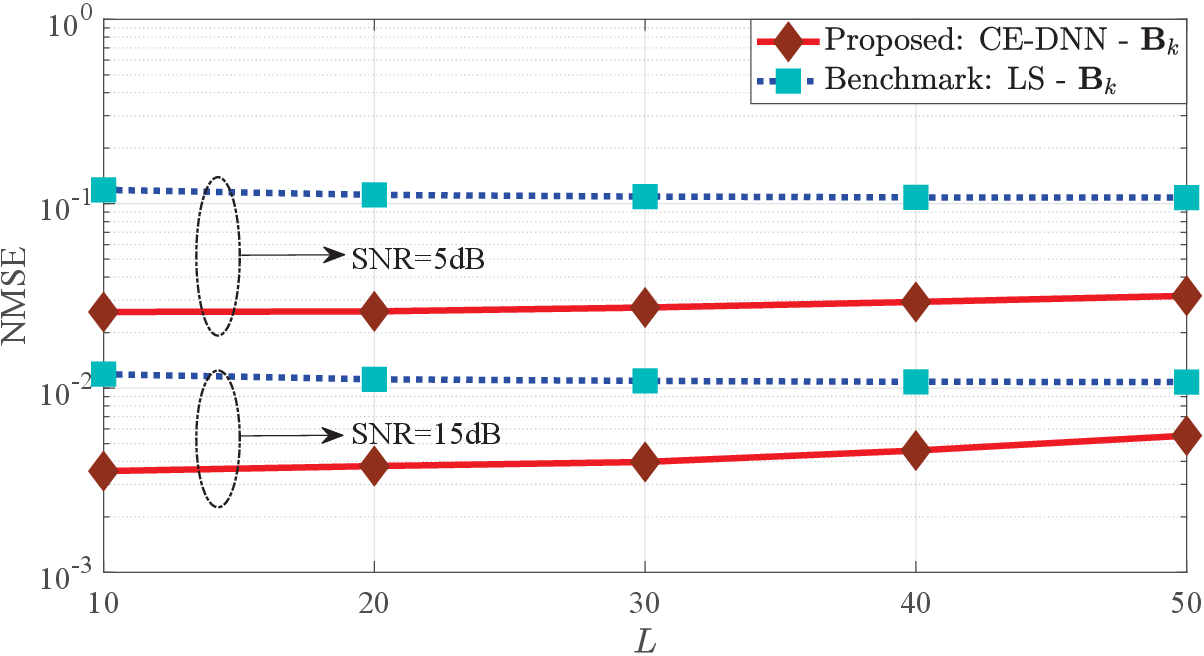}
\caption{NMSE of communication channel estimation versus $L$ \textcolor{black}{at} $M=4$ and $K=3$ \textcolor{black}{for} $\text{SNR}=5\,\rm dB$ \textcolor{black}{and} $15\,\rm dB$. }
\label{fig:NMSE_L}
\end{figure}

Since $L$ directly affects the dimension of the communication channel $\bfB_k$, the impact of \textcolor{black}{varying} $L$ on the estimation performance is assessed in Fig. \ref{fig:NMSE_L}.
The SNR in the offline training and online testing phases is fixed to $5\,\rm dB$ and $15\,\rm dB$ \textcolor{black}{for CE-DNN}.
Obviously, the proposed approach provides significant performance improvement compared with the benchmark \textcolor{black}{scheme for} different \textcolor{black}{$L$ values} and SNR conditions.
However, one can note that the NMSE slightly increases as $L$ increases.
This may lie in the fact that the mapping of the input-output pair $(\bfR^{U_k},\bfH^{U_k})$ is more \textcolor{black}{difficult} to be learned by the CE-DNN when \textcolor{black}{the} channel dimension increases, thus, affecting the accuracy of the estimation performance.

\begin{figure}
\centering
\includegraphics[width=3.7in]{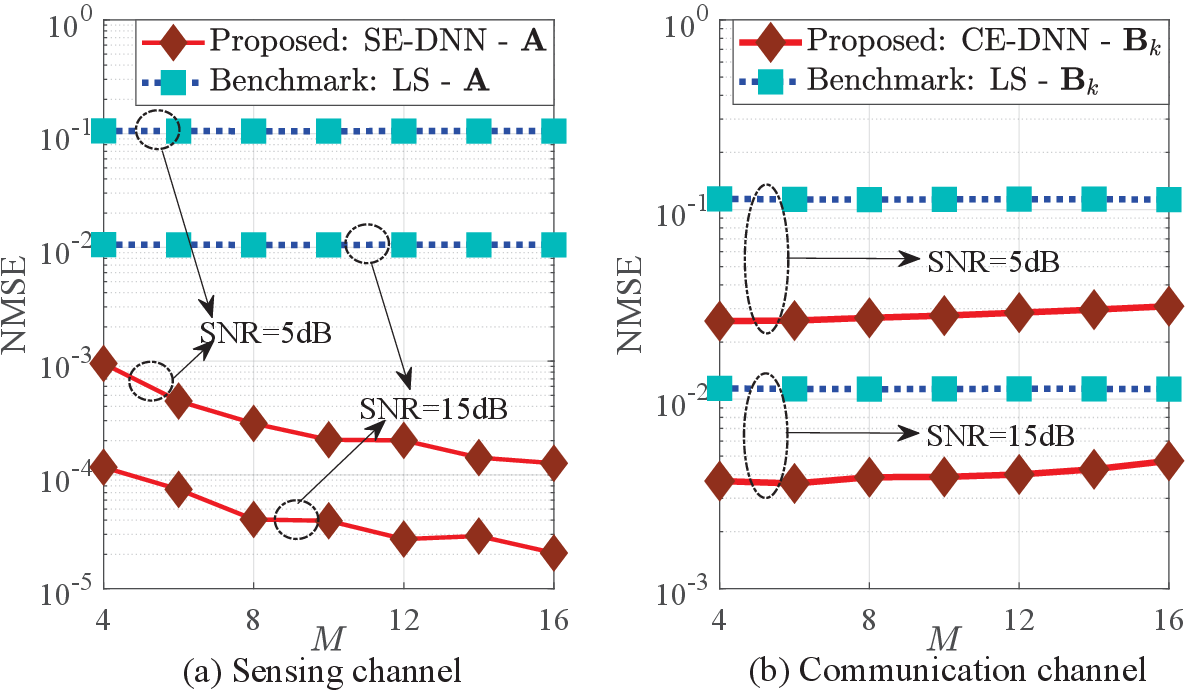}
\caption{NMSE of S\&C channel estimation versus $M$ \textcolor{black}{at} $L=15$ and $K=3$ \textcolor{black}{for} $\text{SNR}=5\,\rm dB$ \textcolor{black}{and} $15\,\rm dB$. }
\label{fig:NMSE_M}
\end{figure}

Fig. \ref{fig:NMSE_M} investigates the impact of increasing $M$ on the NMSE performance.
The SNR setup is the same as \textcolor{black}{for} Fig. \ref{fig:NMSE_L} \textcolor{black}{with} $L=15$.
For the estimation performance of $\bfA$ in \textcolor{black}{Fig. \ref{fig:NMSE_M}(a)}, the NMSE of the proposed approach decreases as $M$ increases and outperforms the benchmark \textcolor{black}{scheme} under different SNR conditions.
The reason is that the proposed SE-DNN can exploit more distinguishable features of $\bfA$ to enhance the estimation accuracy when the channel dimension becomes larger.
\textcolor{black}{Furthermore, the NMSE performance of $\bfB_k$ is depicted in Fig. \ref{fig:NMSE_M}(b).
Similar to the findings of Fig. \ref{fig:NMSE_L}, the NMSE of $\bfB_k$} \textcolor{black}{provided by the proposed approach is superior to that of} \textcolor{black}{the benchmark scheme for different $M$ values and SNR conditions,} \textcolor{black}{and} \textcolor{black}{it slightly increases as $M$ increases.}

\section{\textcolor{black}{Conclusion}}\label{sec:Conclusion}
In this paper, the channel estimation problem for an IRS-assisted ISAC system has been investigated.
To estimate the S\&C channels effectively, a DL estimation framework realized by the SE-DNN and CE-DNN has been proposed, along with \textcolor{black}{the} input-output pairs design.
Numerical results \textcolor{black}{have} shown that under different SNR conditions, the proposed approach possesses superior generalization ability and significantly improves the NMSE performance compared to the benchmark \textcolor{black}{scheme}.
Particularly, the SNR improvements of the proposed approach at $\text{NMSE}=10^{-1}$ are up to $15\,\rm dB$ for the sensing channel estimation, while $5\,\rm dB$ for the communication one.
Furthermore, the proposed approach has been evaluated under a wide range of channel dimensions and \textcolor{black}{results} revealed \textcolor{black}{a} considerable NMSE performance improvement over the benchmark \textcolor{black}{scheme}.

\bibliographystyle{IEEEtran}

\bibliography{Reference_ISAC}

\end{document}